\documentclass[12pt]{article}
\usepackage{graphics}
\usepackage{epsfig}
\usepackage{color}
\newcommand{\rreeG}{$\gamma\gamma\to e^+e^-G_n~$}
\begin{document}
\title{ Associated production of graviton with $e^+e^-$ pair
 via photon-photon collisions at a linear collider
\footnote{Supported by National Natural Science Foundation of
China.}} \vspace{3mm}
\author{{Zhou Ya-Jin, Ma Wen-Gan, Han Liang and Zhang Ren-You }\\
{\small Department of Modern Physics, University of Science and Technology}\\
{\small of China (USTC), Hefei, Anhui 230026, P.R.China}}
\date{}
\maketitle \vskip 12mm
\begin{abstract}
We investigate the process \rreeG at the future International
Linear Collider(ILC), where $G_n$ is the Kaluza-Klein graviton in
the Large Extra Dimension Model. When the fundamental energy scale
is of a few ${\rm TeV}$, the cross section of this process can reach
several hundred $fb$ at a photon-photon collider with
$\sqrt{s}=500 \sim 1000~GeV$, and the cross section in J=2
polarized photon collision mode is much larger than that in J=0
polarized photon collision mode. We present strategies to
distinguish the graviton signal from numerous $SM$ backgrounds,
and find that the graviton signal with extra dimensions $\delta=3$
can be detected when ${\rm M_S} \le 2.67(1.40)~{\rm TeV}$ and
$\gamma \gamma$ c.m.s. energy $\sqrt{s}=1000(500)~{\rm GeV}$ in
unpolarized photon collision mode, while the detecting upper limit
can be increased to 2.79(1.44) ${\rm TeV}$ in $+ -$
($\lambda_1=1$, $\lambda_2=-1$) polarized photon collision
mode(with photon polarization efficiency $P_{\gamma}=0.9$).
\end{abstract}
 \vfill \eject \baselineskip=0.32in
\newcommand{ \slashchar }[1]{\setbox0=\hbox{$#1$}   
   \dimen0=\wd0                                     
   \setbox1=\hbox{/} \dimen1=\wd1                   
   \ifdim\dimen0>\dimen1                            
      \rlap{\hbox to \dimen0{\hfil/\hfil}}          
      #1                                            
   \else                                            
      \rlap{\hbox to \dimen1{\hfil$#1$\hfil}}       
      /                                             
   \fi}                                             %
\renewcommand{\theequation}{\arabic{section}.\arabic{equation}}
\renewcommand{\thesection}{\Roman{section}}
\newcommand{\nb}{\nonumber}
\makeatletter      
\@addtoreset{equation}{section}
\makeatother       
\section{Introduction}
\par
The hierarchy problem of the standard model($SM$) strongly
suggests new physics at TeV scale, and the idea of extra
dimensions($ED$) \cite{ADD,Lykken:1996fj, Witten:1996mz,
Horava:1995qa, Antoniadis:1990ew, Randall:1999ee} might provide a
solution to this problem. The large extra dimension model($LED$,
also called $ADD$ model)\cite{ADD} is the most promising one among
the various extra dimension models. It introduces a fundamental
scale ${\rm M_S}$ in D (D=4+$\delta$) dimension, which is at the
TeV scale, to unify the gravitational and gauge interactions. The
usual Plank scale ${\rm M_P=1/\sqrt{G_N} \sim 1.22 \times
10^{19}~GeV}$ (where $G_N$ is Newton's constant) is related to
${\rm M_S}$ through
\begin{equation}
\label{Mp} {\rm M_P^2 \sim M_S^{2+\delta}~R^{\delta}}
\end{equation}
where $\delta$ is the number of extra dimensions, and $R/2\pi$ is
the radius of the compactified space. The fundamental scale ${\rm
M_S}$ can be at TeV scale if R is large enough, which is at the
same order with electro-weak scale, thus the hierarchy problem is
settled naturally.
\par
From Eq.(\ref{Mp}) we can estimate the value of $R$. If we set
${\rm M_S=1~TeV}$ and $\delta=1$, we have $R \sim 10^{13} cm$,
which is obviously ruled out since it would modify Newton's law of
gravity at solar-system distances. For $\delta = 2$, there exists
$R \sim 1mm$. The latest torsion-balance experiments predict that
an extra dimension must have a size $R \leq 44 \mu
m$\cite{Kapner:2006si}, so $\delta = 2 $ must be ruled out too.
When $\delta \geq 3$, where $R \sim 1nm$, it is possible to detect
graviton signal at high energy colliders.
\par
The large extra dimension model becomes an attractive extension of
the $SM$ because of its possible testable consequences. As
Arkani-Hamed, Dimopoulos, and Dvali\cite{ADD} proposed, the $SM$
particles exist in the usual (3+1)-dimensional space, while
gravitons can propagate in a higher-dimensional space. The picture
of a massless graviton propagating in D-dimensions is equal to the
picture that numerous massive Kaluza-Klein gravitons propagating
in 4 dimensions. So we can expect that even though the
gravitational interactions in the 4 space-time dimensions are
suppressed by a factor of $1/{\rm M_P}$, it can be compensated by
these numerous KK-states. So in either the case that real graviton
emission or the case virtual graviton exchange, it is shown
\cite{Gian,Than} that, after summing over the KK-states, the Plank
mass ${\rm M_P}$ cancels out of the cross section, and we can
obtain an interaction strength comparable to the electroweak
strength.
\par
The CERN Large Hadron Collider(LHC) and the planned International
Linear Collider(ILC) both provide ideal grounds for testing $SM$
and probing possible physics beyond $SM$. However, the ILC has
more advantage in testing extra dimensions. Even though the LHC
and ILC have comparable search reaches for the direct KK graviton
production, the LHC is hampered by theoretical ambiguities due to
a break-down of the effective theory when the parton-level
center-of-mass energy exceeds ${\rm M_S}$\cite{Weiglein:2004hn}.
Furthermore, the ILC has cleaner environment than LHC, so it's
much easier to separate the ED signals. In the first stage of the
ILC, the center-of-mass energy will reach $500~GeV$ and the
luminosity, ${\cal L}$, will be 500 $fb^{-1}$ for the fist four
years running. The second phase foresees an energy upgrade to
about 1 TeV and a luminosity to one ${\rm ab^{-1}}$ in $3-4$ years
running. ILC can also be operated in $\gamma\gamma$ and $e\gamma$
modes, where high energy photon beams can be obtained and easily
polarized via laser back-scattering of the $e^+e^-$ beams.
\par
Since gravitons interact with detectors weakly, they are not
detectable and will give rise to missing energy, so the suggested
graviton signal at LC would be associated production of graviton
with $SM$ particles. The most frequently discussed processes at LC
are associated production of graviton with a photon
($e^+e^-\to\gamma G_n$) \cite{Gian,Than,mirabelli,9909364}, a
fermion pair ($e^+e^-\to e^+e^-G_n$) \cite{Dutta, Atwood},  a Z
boson ($e^+e^-\to Z G_n$)\cite{Cheung}, or a fermion at $e\gamma$
mode ($e\gamma \to f G_n$)\cite{Atwood}. Generally $e^+e^-$
collider has the advantage that the luminosity is higher than
$\gamma\gamma$ collider, for example, ${\cal L}_{\gamma\gamma}
\sim 0.15-0.2 ~{\cal L}_{e^+e^-}$ or even $ 0.3-0.5 ~{\cal
L}_{e^+e^-}$(through reducing emittance in the damping
rings)\cite{Telnov}, but the polarization technique for photon is
much simpler than electron. And through calculation we find that
the W-induced $SM$ background can be reduced after polarization.
Furthermore, $\gamma\gamma \to l^+l^-$ ($l=e,\mu$) is the best
process for the measurement of the $\gamma\gamma$
luminosity\cite{luminosity}, so it is convenient to select events
with missing energy from these beam calibration processes. For
these reasons we studied the associated production of graviton
with an $e^+e^-$ pair at a LC in $\gamma\gamma$ collision mode,
i.e., the process \rreeG. The paper is arranged as follows: in
section II, we present the analytical and numerical calculation of
the cross section. The signal analysis and background elimination
strategies are given in section III. Finally, a short summary is
given.

\par
\section{Cross Section Calculations}
\subsection{Analytical Calculations}
\par
\begin{figure}[htbp]
\scalebox{0.9}[0.9]{\includegraphics*[120pt,580pt][585pt,702pt]{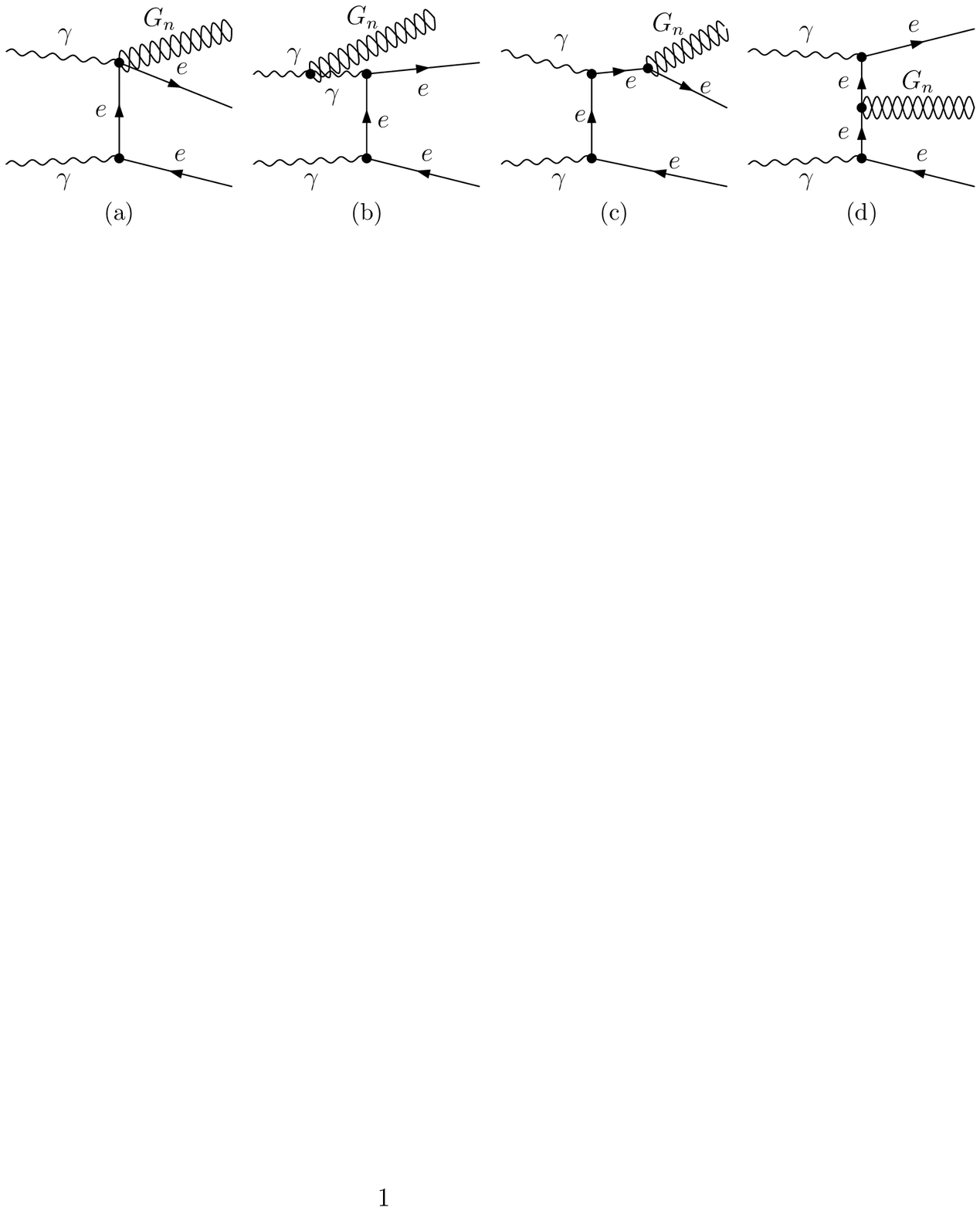}}
\caption{\label{feyn} Representative diagrams for the process
\rreeG} \label{Fig1}
\end{figure}

We denote the process of a graviton production associated with
$e^+e^-$ pair as:
\begin{eqnarray}
\gamma(p_1,\lambda_1,\nu_1)\gamma(p_2,\lambda_2,\nu_2) \to
G_n(k_3,\lambda_s,\mu_1,\mu_2)e^-(k_4)e^+(k_5)
\end{eqnarray}
where $p_i$ and $k_i$ are the momenta of the incoming photons and
outgoing particles respectively, $\lambda_{1,2}$, $\lambda_s$ are
the polarizations of incoming photons and final graviton, and
$\nu_i$, $\mu_i$ are the Lorentz indices of the photons and
graviton respectively. There are 14 Feynman diagrams contributing
to this process at the tree-level, which are representatively
shown in Fig.\ref{Fig1}. The possible corresponding Feynman
diagrams created by exchanging the initial photons and the
graviton radiated from final positron also involved in our
calculation. Fig.\ref{Fig1}(a) gives the Feynman diagram where a
graviton emitted from a vertex, and there are four such kind of
diagrams involved. Fig.\ref{Fig1}(b) represents a graviton emitted
from one of the initial photons via triple vertex, and there are 4
such kind of diagrams. Fig.\ref{Fig1}(c) represents a graviton
emission from the final electrons/positron, and there are also 4
diagrams included. Fig.\ref{Fig1}(d) shows the diagram that a
graviton emission from the electron propagator, and there are two
such diagrams. The Feynman diagrams mediated by a graviton is not
figured in Fig.\ref{Fig1} and also not included in our calculation
because of their neglectable contributions within the energy
regions considered in this paper.
\par
\par
In our calculation we consider both the spin-0 and spin-2 graviton
emission processes, and find that in the case of scalar graviton
emission, only the electron-mass dependent terms give
contributions to the amplitude, just like in the case $f\bar{f}
\to V+ G_n$, which was mentioned in Ref.\cite{Than}. So we are
only interested in the spin-2 component of the Kaluza-Klein(KK)
states. We use the Feynman rules presented in
Refs.\cite{Gian,Than} to calculate the amplitude of process
\rreeG.
\par
The gravitational coupling $\kappa \equiv \sqrt{16 \pi G_N}$ can
be expressed in terms of the fundamental scale ${\rm M_S}$ and the
size of the compactified space R by
\begin{eqnarray}
  \kappa^2 R^{\delta} = 16 \pi (4 \pi)^{\delta/2} \Gamma(\delta/2) M_S^{-(\delta+2)}
\end{eqnarray}
Here we give the amplitudes for the representative diagrams shown
in Fig.\ref{feyn}(a-d), separately.
\begin{eqnarray} \nonumber
{\cal M}_a&=&\frac{ie^2\kappa}{4} \frac{1}{(k_5-p_2)^2}
\epsilon_{\nu_1}(p_1)\epsilon_{\nu_2}(p_2)\epsilon_{\mu_1\mu_2}(k_3)
\bar{u}(k_4)
(\gamma_{\mu_2}\eta_{\mu_1\nu_1}+\gamma_{\mu_1}\eta_{\mu_2\nu_1}
-2\gamma_{\nu_1}\eta_{\mu_1 \mu_2} )\\ && (\slashchar{p_2}
-\slashchar{k_5}) \gamma_{\nu_2}v(k_5), \\ \nb
{\cal M}_b&=&\frac{ie^2\kappa}{2}\frac{1}{(p_1-k_3)^2}\frac{1}{(k_5-p_2)^2}
             \epsilon_{\nu_1}(p_1)\epsilon_{\nu_2}(p_2)\epsilon_{\mu_1\mu_2}(k_3)
             \bar{u}(k_4)\gamma_{\alpha}
             (\slashchar{p_2}-\slashchar{k_5})\gamma_{\nu_2}v(k_5) \\ \nb
          && \bigg{\{} \frac{1}{2}\eta_{\mu_{1}\mu_{2}}
             \bigg{[} -p_{1\alpha}(p_1-k_3)_{\nu_{1}}+p_1\cdot (p_1-k_3) \eta_{\nu_{1}\alpha} \bigg{]}
             - \eta_{\nu_{1}\alpha}p_{1\mu_{1}}(p_1-k_3)_{\mu_{2}}  \\ \nb
      && + \eta_{\mu_{1}\nu_{1}}\bigg{[} -p_1\cdot (p_1-k_3) \eta_{\mu_{2}\alpha}+p_{1\alpha}(p_1-k_3)_{\mu_{2}}
              \bigg{]}-\eta_{\mu_{1}\alpha}p_{1\mu_{2}}(p_1-k_3)_{\nu_{1}}  \\ \nb
      && + \frac{1}{2}\eta_{\mu_{2}\mu_{1}} \bigg{[} -p_{1\alpha}(p_1-k_3)_{\nu_{1}}+p_1\cdot (p_1-k_3)
          \eta_{\nu_{1}\alpha} \bigg{]} - \eta_{\nu_{1}\alpha}p_{1\mu_{2}}(p_1-k_3)_{\mu_{1}}   \\
          && + \eta_{\mu_{2}\nu_{1}} \bigg{[} -p_1\cdot (p_1-k_3)
             \eta_{\mu_{1}\alpha}+p_{1\alpha}(p_1-k_3)_{\mu_{1}} \bigg{]}
             +\eta_{\mu_{2}\alpha}p_{1\mu_{1}}(p_1-k_3)_{\nu_{1}}  \bigg{\} },  \\ \nb
{\cal M}_c&=&-\frac{ie^2\kappa}{8}
             \frac{1}{(k_3+k_4)^2}\frac{1}{(p_2-k_5)^2}
             \epsilon_{\nu_1}(p_1)\epsilon_{\nu_2}(p_2)\epsilon_{\mu_1\mu_2}(k_3)
             \bar{u}(k_4) \bigg{[} \gamma_{\mu_2}(k_3+ 2k_4)_{\mu_1}+ \\
          && \gamma_{\mu_1}(k_3+2k_4)_{\mu_2}-2\eta_{\mu_1
             \mu_2}(\slashchar{k_3}+2\slashchar{k_4}) \bigg{]}
             (\slashchar{k_3}+\slashchar{k_4})\gamma_{\nu_1}(\slashchar{p_2}-
             \slashchar{k_5})\gamma_{\nu_2}v(k_5),  \\ \nb
{\cal M}_d&=&-\frac{ie^2\kappa}{8} \frac{1}{(p_1-k_4)^2}\frac{1}{(p_2-k_5)^2}
              \epsilon_{\nu_1}(p_1)\epsilon_{\nu_2}(p_2)\epsilon_{\mu_1\mu_2}(k_3)
              \bar{u}(k_4) \gamma_{\nu_1} (\slashchar{k_4}- \slashchar{p_1})
              \bigg{[} \gamma_{\mu_2}(k_3  \\ \nb
      &&  +2k_4- 2p_1)_{\mu_1}+\gamma_{\mu_1}(k_3+2k_4-
              2p_1)_{\mu_2}-2\eta_{\mu_1\mu_2}(\slashchar{k_3}+2\slashchar{k_4}-2\slashchar{p_1})\bigg{]}
              (\slashchar{p_2}-\slashchar{k_5}) \\
      && \gamma_{\nu_2}v(k_5).
\end{eqnarray}
The amplitudes for other diagrams can be easily obtained by
changing the corresponding momenta in these expressions.

\par
The spin-averaged amplitude squared for the process is expressed
as follow:
\begin{eqnarray}
\overline{\sum_{spins}}\vert{\cal
M}\vert^2=\frac{1}{4}\sum_{spins}\left(\sum_{i=1}^{14}{\cal
M}_{i}\right)^{\dag}\left(\sum_{i=1}^{14}{{\cal M}_{i}}\right)
\end{eqnarray}
where ${\cal M}_i$ represents the amplitude for the $i$th Feynman
diagram. The bar over summation means taking average over initial
photon spin states. By taking the summation over the polarizations
of the spin-2 graviton tensors wave functions, we have
\cite{Gian,Than}:
\begin{eqnarray}
\sum_{\lambda_s=1}^5 \epsilon_{\mu\nu}(k,\lambda_s)\epsilon^*_{\alpha\beta}(k,\lambda_s)=
  P_{\mu\nu \alpha\beta}(k) .
\end{eqnarray}
where $P_{\mu\nu\alpha\beta}$ is:
\begin{eqnarray}
\label{Pmunu}\nb
P_{\mu\nu\alpha\beta} & = &\frac{1}{2}\left( \eta_{\mu\alpha}\eta_{\nu\beta}
 +\eta_{\mu\beta}\eta_{\nu\alpha}
  -\eta_{\mu\nu}\eta_{\alpha\beta}\right)
\label{propg}
\\ \nonumber
 & & -\frac{1}{2m^2}\left(
\eta_{\mu\alpha}{k_\nu k_\beta}
 +\eta_{\nu\beta}{k_\mu k_\alpha}+
  \eta_{\mu\beta}{k_\nu k_\alpha} +\eta_{\nu\alpha}
  {k_\mu k_\beta}\right) \\
 & & +\frac{1}{6}
\left( \eta_{\mu\nu}
    +\frac{2}{m^2}    k_\mu k_\nu  \right)
\left( \eta_{\alpha\beta}
    +\frac{2}{m^2}    k_\alpha k_\beta  \right) .
\end{eqnarray}
\par
In practical experiments, the contributions of the different
Kaluza-Klein modes have to be summed up. For not too large extra
dimensions $\delta$, the mass spacing of these KK-states is much
smaller than the physical scale, so it is convenient to replaced
the summation over the KK-states by a continuous integration:
\begin{eqnarray}
\label{replace} \sigma=\sum_{n} \sigma_m \to \int_{0}^{\sqrt{s}}
\rho(m)~ \sigma_m~ dm,
\end{eqnarray}
where $\rho(m)$ is the density of states, which is
\begin{eqnarray}
\label{density} \rho(m) = \frac{2
R^{\delta}m^{\delta-1}}{(4\pi)^{\delta/2}\Gamma(\delta/2)} =
\frac{32 \pi m^{\delta-1}}{\kappa^2 M_S^{\delta +2}}.
\end{eqnarray}
$\sigma_m$ in Eq.(\ref{replace}) is the cross section for a
definite KK-state, and it can be expressed as the integration over
the phase space of three-body final states:
\begin{eqnarray}
\sigma_{m}&=&\frac{(2\pi)^4 }{4|\vec p_1|\sqrt{s}}\int d\Gamma_3
\overline{\sum_{spins}}|{\cal M}|^2.
\end{eqnarray}
The integration is performed over the three-body phase space of
final particles $e^+e^-G_n$. The phase-space element $d\Gamma_3$
is defined by
\begin{eqnarray}
{d\Gamma_3}=\delta^{(4)} \left( p_1+p_2-\sum_{i=3}^5 k_i \right)
\prod_{j=3}^5 \frac{d^3 \textbf{\textsl{k}}_j}{(2 \pi)^3 2 E_j}.
\end{eqnarray}
In the process of numerical calculation, the integration over the
mass of the KK-states and over the phase space can be done at the
same time.

\par
\subsection{Numerical Results}
\par
In this subsection we present the numerical results of the total
cross section for the process \rreeG. The value of the fine
structure constant $\alpha$ is taken as $1/128$ \cite{mass}. In
the calculations for this process with TeV scale colliding energy
we ignore the masses of electron and positron. To remove the
singularities which arise when the final massless
electron/positron is collinear with the photon beam, we set a
small cut on the angle between electron/positron and one of the
incoming photons, which is $2^\circ < \theta_{e\gamma}<
178^\circ$.
\par
The incoming $\gamma$ beams have five polarization modes: $+~+$,
$+~-$, $-~+$, $-~-$ and unpolarized collision modes, for example,
the notation of $+~-$ represents helicities of the two initial
photons being $\lambda_1=1$ and $\lambda_2=-1$. In Table 1 we give
the total cross sections of the graviton emitting process
accompanied with an $e^+e^-$ pair considered in this paper, with
different numbers of extra dimensions, $\gamma-\gamma$ c.m.s.
energy, and photon polarization modes. The polarization efficiency
of photon $P_{\gamma}$($P_{\gamma}\equiv\frac{N_+ - N_-}{N_+ +
N_-}$) is assumed to be 0.9. Since the cross sections of the $+~-$
and $-~+$ photon polarizations (i.e., J=2) are equal, and also the
cross sections of the $+~+$ and $-~-$ photon polarizations (i.e.,
J=0) are the same, we only give the total cross sections in three
cases: $+~-$, $+~+$ and unpolarized photons.
\begin{table}[htb]
\centering \caption{Total cross sections for the process \rreeG,
with and without photon polarization. ${\rm M_S}$ is set to be 1
TeV, the polarization efficiency $P_{\gamma}=0.9$, and the cross
sections are in $fb$.} \vskip 0.3cm
\begin{tabular}{|l|c|c|c|c|c|c|}
\hline $\sqrt{s}$ [GeV]&& $\delta =3$ & $\delta=4$
& $\delta =5 $ & $\delta =6$\\
\hline
\hline
        & unpol.   &46.46 &13.92 &4.692 &1.700 \\
~~~500  & $+~-$    &60.01 &19.35 &6.853 &2.576 \\
        & $+~+$    &32.91 &8.493 &2.532 &0.821 \\
\hline \hline
        & unpol.   &371.7 &222.7 &150.1 &108.8 \\
~~~1000 & $+~-$    &480.8 &309.6 &219.3 &164.9 \\
        & $+~+$    &262.6 &135.8 &80.93 &52.75 \\
\hline
\end{tabular}
\end{table}
\par
\begin{figure}[htb]
\vskip -3 cm
\includegraphics[scale=0.45,bb=30 30 520 594]{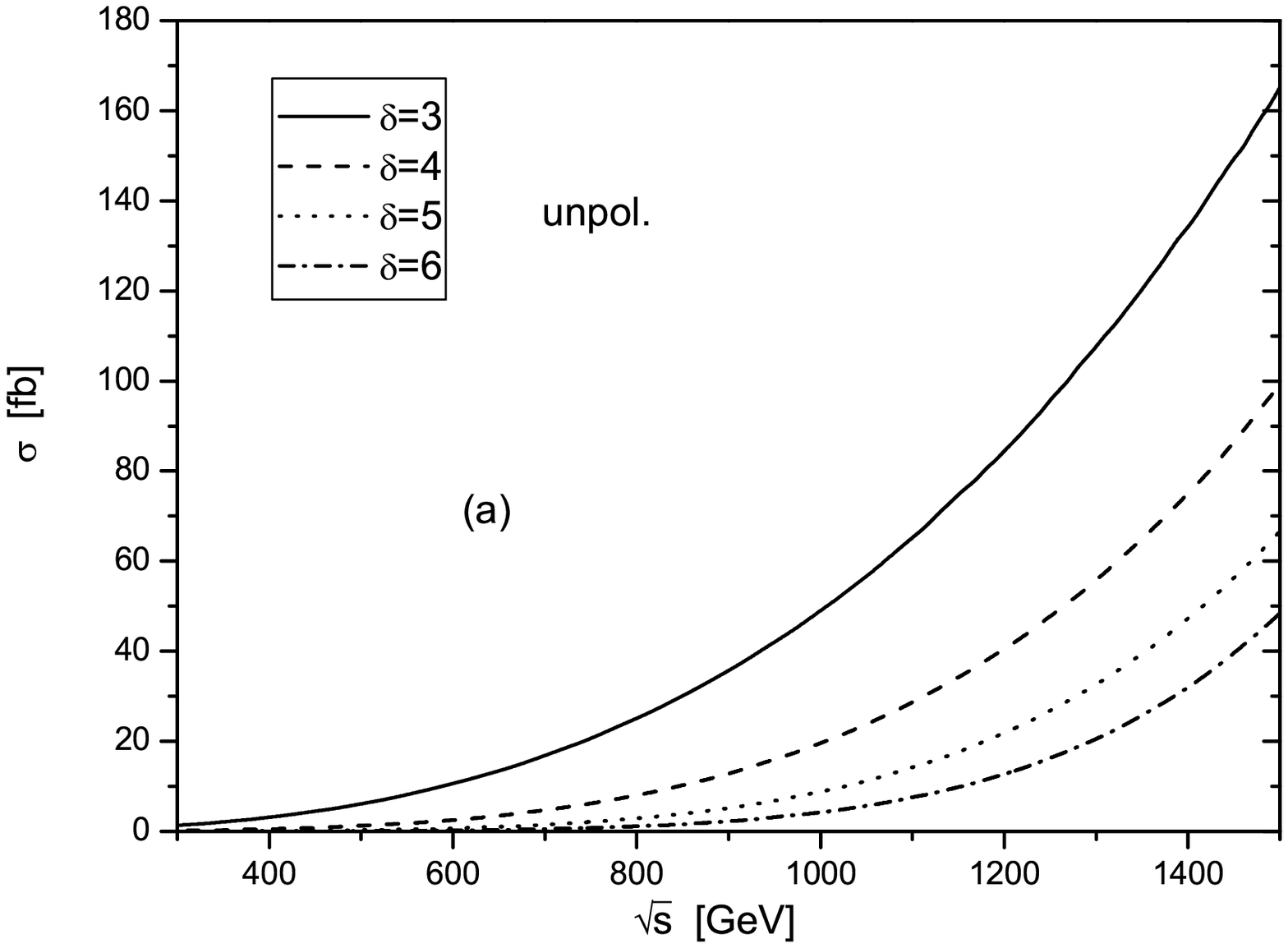}
\includegraphics[scale=0.45,bb=30 30 520 390]{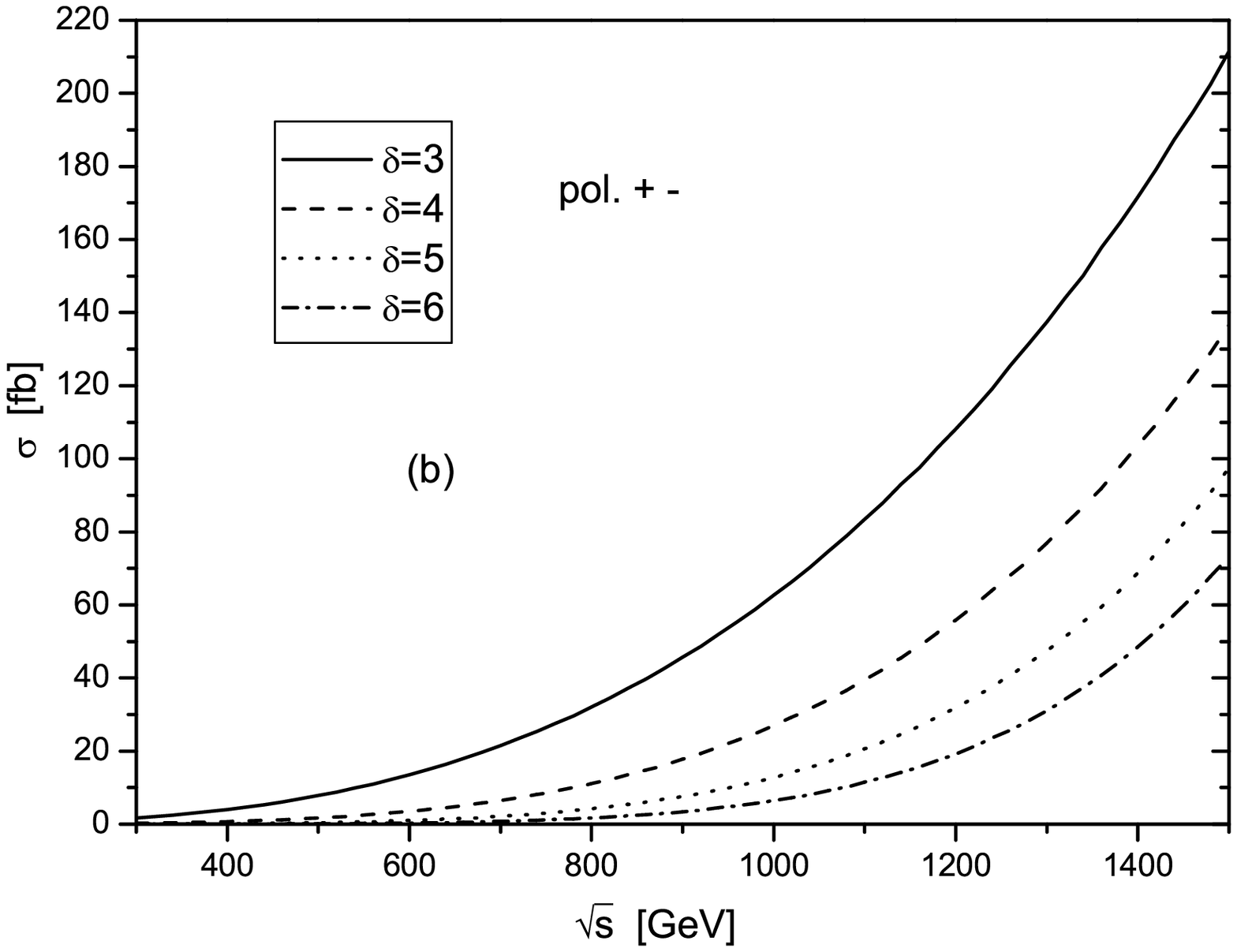}
\caption{\label{cross-sqrts}The cross section for process \rreeG
as the functions of $\gamma \gamma$ colliding energy $\sqrt{s}$,
with ${\rm M_S = 1.5~TeV}$. (a) for the unpolarized photon-photon
collision, (b) for +~- polarized photon-photon collision with $P_{\gamma}=0.9$.}
\end{figure}

From Table 1 we can see that the cross section can reach several
hundred $fb$ when the $\gamma-\gamma$ c.m.s. energy is 1 TeV. The
larger the number of extra dimensions is, the smaller the cross
section becomes. It is obvious that the cross sections in the case
with +~- polarized incoming photons are much larger than those in
the case in +~+ polarized photon-photon collision. This is because
the spin of the emitted graviton is 2, so it is much easier for
J=2 initial states to generate a spin-2 graviton. This feature
becomes more evident when the number of extra dimensions becomes
larger. When $\delta=6$, the cross section with the +~- polarized
photons can reach three times of that with the +~+ polarized
photons. We also find that the cross sections at $\sqrt{s}=1~{\rm
TeV}$ are much larger than those at $\sqrt{s}=500~GeV$. To show
the relationship between the cross section and the $\gamma-\gamma$
c.m.s energy more clearly, we depict two curves for the production
rate of the process \rreeG as the function of $\sqrt{s}$ in
Fig.\ref{cross-sqrts}, with ${\rm M_S=1.5~TeV}$ and the incoming
photons being unpolarized and +~- polarized($P_{\gamma}$=0.9),
respectively. Since the perturbative theory is only applicable
when $\sqrt{s} \leq {\rm M_S}$, we take the $\sqrt{s} < 1.5~{\rm
TeV}$ in Fig.\ref{cross-sqrts}. The figure shows that the cross
sections go up quickly with the increment of $\sqrt{s}$, because
there are more KK-states contribute to the cross section. And the
cross section is rather small at low $\sqrt{s}$ because of phase
space suppression. At the same time we can see that the $+~-$
polarization photon beams can strongly enhance the cross section.

\par
\section{Signal analysis}
Since graviton interacts with materials weakly, an emitted
graviton cannot be detected in experiment. Therefore, in the
measurement of the process \rreeG  the existence of a graviton
manifests as the phenomenon of missing energy. The signature for
this process is:
\begin{eqnarray}
\gamma\gamma \to e^+ e^- + missing~energy.
\end{eqnarray}
So the processes of the form $\gamma\gamma \to e^+e^-
(neutrinos)$, where the neutrinos can be of any generation, are
$SM$ background which can effect the discrimination of the
graviton. The main $SM$ background processes at the lowest order
for the signal of \rreeG are:
\begin{eqnarray}
\label{rree}  &&\gamma\gamma \to e^+e^-  \\
\label{rreeZ} &&\gamma\gamma \to e^+e^-Z \to e^+e^-(\nu\bar{\nu}) \\
\label{rrww}  &&\gamma\gamma \to W^+W^- \to (e^+ \nu_e)(e^- \bar{\nu_e}) \\
\label{rrtt}  &&\gamma\gamma \to \tau ^+ \tau ^-
    \to (e^+ \nu_e \bar{\nu_{\tau}})(e^- \bar{\nu_e} \nu_{\tau})
\end{eqnarray}
\par
All these processes contribute a formidable background to the
graviton signal of \rreeG. However, a reasonable set of kinematic
cuts enable us to distinguish the suggested signal from the
backgrounds. Firstly, the electron-positron pair in the process
(\ref{rree}) are collinear, so we can remove the $\gamma\gamma \to
e^-e^+$ background totally by putting a cut on the angle between
the final electron and positron. The contributions from
$\gamma\gamma\to e^+e^-Z$ can be expressed as:
\begin{eqnarray}\nb
\sigma _{e^+e^-Z}
&=& \sigma (\gamma\gamma  \to e^+e^-Z) \times
 Br(Z \to \nu\bar{\nu})  \\
&=& \sigma (\gamma\gamma  \to e^+e^-Z) \times 20.0\%  \\
&=& \cases { 10.65~fb ~~~~~(for~\sqrt{s}=500~GeV); \cr
              4.17~fb ~~~~~(for~\sqrt{s}=1000~GeV). }
\end{eqnarray}

\par
The primary dominant backgrounds should be the $\gamma\gamma \to
W^+W^-$ and $\gamma\gamma \to \tau ^+ \tau ^-$ processes, which
are called as WW- and $\tau\tau$-background respectively in the
following discussion. Their contributions are given by
\begin{eqnarray}\nb
\sigma _{WW}
&=& \sigma (\gamma\gamma \to W^+W^-) \times
\left (Br(W \to e\nu_e)\right )^2 \\
&=& \sigma (\gamma\gamma \to W^+W^-) \times (10.75\%)^2 \\
  &=& \cases {1011~fb ~~~~~(for~\sqrt{s}=500~GeV) \cr
              1019~fb ~~~~~(for~\sqrt{s}=1000~GeV).}
\end{eqnarray}
\begin{eqnarray}\nb
\sigma _{\tau\tau}&=& \sigma (\gamma\gamma \to \tau^+\tau^-) \times
\left (Br(\tau \to e\nu_e \nu_{\tau})\right )^2 \\
&=& \sigma (\gamma\gamma \to \tau^+\tau^-) \times (17.84\%)^2 \\
    &=& \cases {243.1~fb ~~~~~(for~\sqrt{s}=500~GeV)  \cr
                62.36~fb ~~~~~(for~\sqrt{s}=1000~GeV). }
\end{eqnarray}

\par
\begin{figure}[htbp]
\centerline{\epsfxsize=16cm \epsfysize=12cm\epsfbox{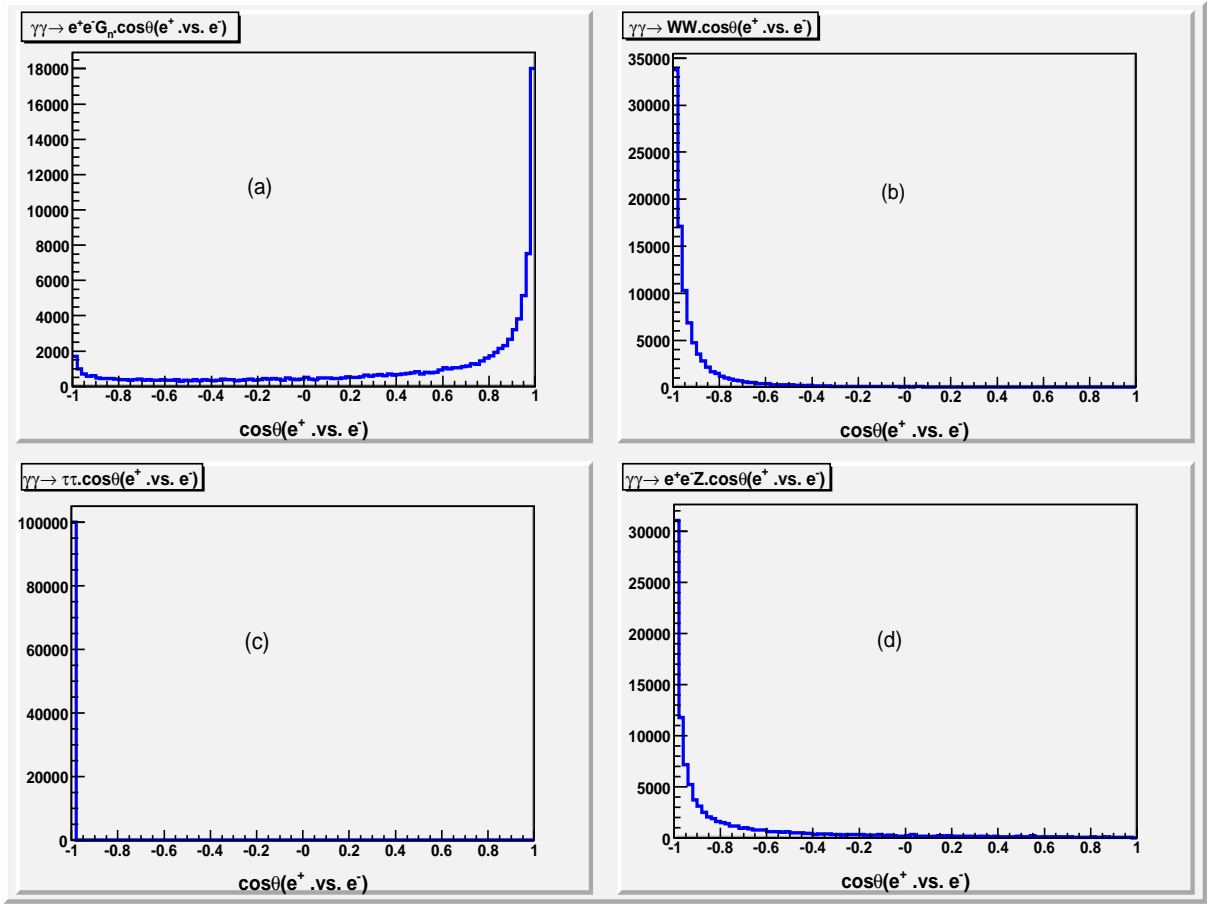} }
\caption{\label{distri} Distributions of the open angel between
the electron and positron for the signal process when extra
dimensions $\delta=3$ and the background processes. The
$\gamma\gamma$ c.m.s. energy is 1 TeV and ${\rm M_S}$ is set to be
1 TeV.}
\end{figure}
\begin{figure}[htbp]
\centerline{\epsfxsize=16cm \epsfysize=6cm\epsfbox{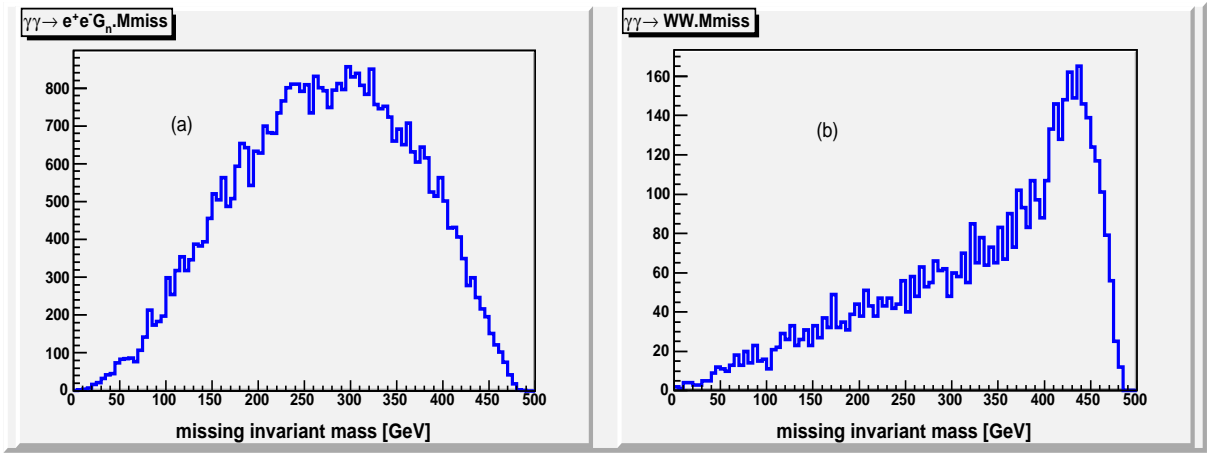}
} \caption{\label{distri-Mmiss} Distributions of the missing
invariant mass of the signal process and the WW-background process
after applying $CUT1$ when extra dimensions $\delta=3$. The
$\gamma\gamma$ c.m.s. energy is $500~GeV$ and ${\rm M_S}$ is set
to be 1 TeV.}
\end{figure}

We developed an event generator program for the process \rreeG and
$\gamma\gamma\to e^+e^-Z$, and the WW- and $\tau\tau$-background
events are generated by adopting Pythia package\cite{pythia}. In
Fig.\ref{distri} we depict the Monte Carlo distributions of the
open angle between electron and positron $\theta_{ee}$ for the
signal process \rreeG and background processes($\gamma\gamma\to
e^+e^-Z$, $\gamma\gamma\to W^+W^-$ and $\gamma\gamma\to \tau\tau$)
separately, with the $\gamma\gamma$ colliding energy
$\sqrt{s}=1~{\rm TeV}$. The $\theta_{ee}$ distributions for the
signal process \rreeG, $\gamma\gamma\to e^+e^-Z$, WW- and
$\tau\tau$-background processes at $\sqrt{s}=500~{\rm GeV}$ have
the similar line-shapes with the corresponding ones at
$\sqrt{s}=1000~{\rm GeV}$ as shown in Fig.\ref{distri}. The
back-to-back feature of the final $e^+e^-$ pair for the main
background processes is shown evidently in Fig.\ref{distri},
especially for the process $\gamma\gamma \to \tau\tau$. If we put
a suitable cut on the open angle between the electron and positron
$\theta_{ee}^{cut}$, it is possible to remove the background
events including also the $\gamma\gamma\to e^+e^-$ process from
the signals of \rreeG  at $\sqrt{s}=1000~{\rm GeV}$. But in the
case of $\sqrt{s}=500~{\rm GeV}$, the $\theta_{ee}^{cut}$ is not
enough to eliminate WW-background process. In
Fig.\ref{distri-Mmiss} we show the simulating distributions of the
missing invariant mass of the signal process \rreeG and the
WW-background process with extra dimensions $\delta=3$,
$\sqrt{s}=500~GeV$ and ${\rm M_S=1~TeV}$ after applying $CUT1$
which is introduced below, and find that an extra missing
invariant mass cut can reduce more WW-background events.
\par
From the above discussion we choose the off-line event selection
criterions as follows:
\begin{enumerate}
\item To take into account the detector acceptance, firstly, we
demand that the angle between electron(positron) and the photon
beam should be in the range
$5^{\circ}<\theta_{e\gamma}<175^{\circ}$. Secondly, the transverse
momentum of the electron(positron) should satisfy $p_T^e >5~{\rm
GeV}$. We also demand that the electron(positron) energy
$E_e>1GeV$. To separate the electron and positron tracks, we
demand that the open angle between electron and positron
$\theta_{ee}$ should be large than $5^{\circ}$. On the other hand,
to eliminate the WW, $\tau\tau$, $\gamma\gamma\to e^+e^-$ and
$e^+e^-Z$ background, we set a more strict cut on $\theta_{ee}$,
i.e., $5^{\circ}<\theta_{ee} <\theta_{ee}^{cut}=90^{\circ}$. We
denote these cuts as $CUT1$ set, which are expressed as
\begin{eqnarray}\nb
5^{\circ}<\theta_{e\gamma}<175^{\circ},~
p_T^e >5~GeV, ~E_e > 1~GeV, ~{\rm and}~
5^{\circ}<\theta_{ee}<\theta_{ee}^{cut}=90^{\circ}
\end{eqnarray}
\item For $\sqrt{s}=500~{\rm GeV}$, a cut on the missing invariant
mass is needed, which is denoted as $CUT2$:
$$M_{miss}<400~GeV$$
\end{enumerate}
\par
Taking the photon integrated luminosity to be ${\cal L}=100^{-1}
fb$, we present in Table 2 the event numbers of the signal process
with $\delta=3$ and ${\rm M_S=1~TeV}$ and the background processes
after each step of cut, the event selection efficiency after cuts,
and the significance of signal over background. Here the event
selection efficiency is defined as the event numbers after cuts
divided by the event numbers before any cut. The significance of
signal over background is defined as
\begin{eqnarray}
SB&=&\frac{N_{signal}}{\sqrt{N_{background}}}
=\frac{\sigma_S^{CUT}\cdot {\cal L}_{\gamma\gamma}}
{\sqrt{\sigma_B^{CUT}\cdot {\cal L}_{\gamma\gamma}}}\\ \nb
&=& \frac{\sigma_S^{CUT}}{\sqrt{\sigma_B^{CUT}}}\cdot\sqrt{{\cal L}_{\gamma\gamma}}
\end{eqnarray}
\begin{table}[htb]
\centering \caption{Event selection on background and
signal($\delta=3$) with unpolarization case.}
\begin{tabular}{|c|c|c|c|c|c|c|c|c|}
\hline
& \multicolumn{4}{|c|}{$\sqrt{s}=500GeV$} & \multicolumn{4}{|c|}{$\sqrt{s}=1000GeV$} \\
\hline
& $e^+e^-G_n$ & WW & $\tau\tau$ &$e^+e^-Z$ & $e^+e^-G_n$ & WW  &$\tau\tau$ &$e^+e^-Z$\\
\hline
N before cut   & 4646 & 101100 & 24310 & 1065 & 37170 & 101900  & 6236 & 417 \\
\hline
N after $CUT1$ & 1805 & 5402   & 0     & 86  & 17790 & 649     & 0    & 31  \\
\hline
N after $CUT2$ & 1616 & 3427   & 0     & 86  & /     & /    & /     & / \\
\hline
efficiency $\epsilon$     & 34.8\% & 3.39\% & 0\% & 8.08\% & 47.9\% & 0.64\% & 0\% & 7.43\% \\
\hline
SB &\multicolumn{4}{|c|}{27.26} & \multicolumn{4}{|c|}{682.2}  \\
\hline
\end{tabular}
\end{table}
\par
Here we have discussed the case of unpolarized photons. In Section
II we show that the cross section for the signal process with
$J=2$ polarized photons is much larger than that with $J=0$
photons. On the contrary, we find that with the incoming photon
polarizations, the cross section for the primary $SM$ background
process $\gamma\gamma \to W^+W^-$ is suppressed in the colliding
case with $J=2$. So it will be much easier to eliminate $SM$
backgrounds in case of $J=2$ collision mode. Using the similar
signal analysis procedure with above, we obtain the data with $+
-$ polarized case(with $P_{\gamma}=0.9$), and list them in Table
3.
\begin{table}[htb]
\centering \caption{Event selection on background and
signal($\delta=3$),with $+-$ polarized photon beams($P_{\gamma}=0.9$).}
\begin{tabular}{|c|c|c|c|c|c|c|c|c|}
\hline
& \multicolumn{4}{|c|}{$\sqrt{s}=500GeV$} & \multicolumn{4}{|c|}{$\sqrt{s}=1000GeV$} \\
\hline
& $e^+e^-G_n$ & WW & $\tau\tau$ &$e^+e^-Z$ & $e^+e^-G_n$ & WW  &$\tau\tau$ &$e^+e^-Z$\\
\hline
N before cut   & 5926 & 96159 & 38271 & 1340 & 47400 & 99065  & 10299 & 480 \\
\hline
N after $CUT1$ & 2104 & 5152   & 0     & 59  & 21524 & 631    & 0    & 19  \\
\hline
N after $CUT2$ & 1782 & 3268   & 0     & 59  & /     & /    & /     & / \\
\hline
efficiency $\epsilon$     & 30.1\% & 3.40\% & 0\% & 4.40\% & 45.4\% & 0.64\% & 0\% & 3.96\% \\
\hline
SB &\multicolumn{4}{|c|}{30.89} & \multicolumn{4}{|c|}{844.2}  \\
\hline
\end{tabular}
\end{table}
\par
Notice that the data in Table 2 and Table 3 are obtained by taking
${\rm M_S=1~TeV}$. The cross section of the signal process is
proportional to $1/M_S^{\delta+2}$(see
Eq.(\ref{replace}-\ref{density})), so the SB value is proportional
to $1/M_S^{\delta+2}$, too. If we suppose that the signature can
be detected only when $SB \ge 5$ in experiment, then we can reach
the conclusion that in the case of $\sqrt{s}=1~{\rm TeV}$,
$\delta=3$ and unpolarized photon beams, graviton signal can be
detected when ${\rm M_S} \le 2.67 ~{\rm TeV}$, while in the case
of $\sqrt{s}=500~{\rm GeV}$, the graviton signal can be detected
only when ${\rm M_S} \le 1.40~{\rm TeV}$. These limits are
increased to 2.79 ${\rm TeV}$(when $\sqrt{s}=1~{\rm TeV}$) and
1.44 ${\rm TeV}$(when $\sqrt{s}=500~{\rm GeV}$) in $+ -$ polarized
photon collision mode with $P_{\gamma}=0.9$, respectively.

\section{Summary}
\par
In this paper we calculate the cross sections for the process
\rreeG in different polarized photon collision modes, and present
some strategies to discriminate the graviton signal from numerous
$SM$ backgrounds.
\par
At the stage of ILC with $\sqrt{s}= 1~{\rm TeV}$, the cross
section for the process \rreeG can reach several hundred $fb$.
Because the spin of the emitted graviton is 2, the $\gamma\gamma$
collision with $J=2$ strongly enhances the production rate of
process \rreeG, especially when the number of extra dimensions is
large. Of course, the cross section increases with the increment
of c.m.s. energy $\sqrt{s}$, due to more KK-states exist which
contribute to the cross section. Another effect of the $LED$ shown
in this paper is that the cross section decreases when the number
of extra dimensions $\delta$ goes up.
\par
For the case of $\delta=3$,
we present some strategies to select graviton signals from the
numerous $SM$ backgrounds. Because most of the $e^+e^-$ pair of
the background processes are back-to-back, taking cut on the open
angle between electron and positron can reduce the backgrounds
efficiently. With our suggested event selection criterions, we can
reach rather high $SB$ value. We conclude that by adopting an
unpolarized $\gamma\gamma$ collision machine with $\sqrt{s}=1~{\rm
TeV}$ in the case of $\delta=3$ and ${\cal L} = 100 fb^{-1}$, the
graviton signal can be detected when ${\rm M_S} \le 2.67 ~{\rm
TeV}$, while in the case of $\sqrt{s}=500~{\rm GeV}$, the graviton
signal can be detected only when ${\rm M_S} \le 1.40~{\rm TeV}$.
If we adopt a $\gamma \gamma$ collider machine in $+ -$ polarized
photon collision mode, the detecting upper limits on the
fundamental scale can be improved up to 2.79 ${\rm TeV}$ when
$\sqrt{s}=1~{\rm TeV}$, and 1.44 ${\rm TeV}$ when
$\sqrt{s}=0.5~{\rm TeV}$.

\par
\vskip 10mm \noindent{\large\bf Acknowledgments:} This work was
supported in part by the National Natural Science Foundation of
China, the Education Ministry of China and a special fund
sponsored by Chinese Academy of Sciences.

\end{document}